\documentclass[12pt]{amsart}
\usepackage{amssymb}
\usepackage{color}
\usepackage{amsmath,epic,curves,amscd}
\usepackage[english]{babel}
\usepackage{graphicx}
\usepackage{comment}
\usepackage{appendix}
\usepackage{mathdots}
\usepackage[all]{xy}
\newtheorem{theorem}{Theorem}

\theoremstyle{definition}

\newcommand\lan{\langle}
\newcommand\ran{\rangle}
\newcommand\tr{{\text{\rm Tr}}\,}
\newcommand\ot{\otimes}

\newcommand\meet{\wedge}
\renewcommand\ker{{\text{\rm Ker}}\,}

\newcommand\rk{{\text{\rm rank}}\,}

\newcommand\SR{{\text{\rm SR}}\,}

\newcommand\HS{{\text{\rm HS}}}

\newcommand\sr{{\text{\rm SR}}\,}
\newcommand\blockpos{{{\mathcal B\mathcal P}}}

\begin{document}
\baselineskip 6.0 truemm
\parindent 1.5 true pc

\title{Global locations of Schmidt number witnesses}

\author{Kyung Hoon Han and Seung-Hyeok Kye}
\address{Kyung Hoon Han, Department of Data Science, The University of Suwon, Gyeonggi-do 445-743, Korea}
\email{kyunghoon.han at gmail.com}
\address{Seung-Hyeok Kye, Department of Mathematics and Institute of Mathematics, Seoul National University, Seoul 151-742, Korea}
\email{kye at snu.ac.kr}
%\thanks{Both KHH and SHK were partially supported by NRF-2020R1A2C1A01004587, Korea}

%%%%%%%%%%%%%%%%%%%%%%%%%%%%%%%%%%%%%%%%%%%%%%%%%%%%%%%%%%%%%%%%%%%%%%%%%%
%                                                                        %
%                              Abstract                                  %
%                                                                        %
%%%%%%%%%%%%%%%%%%%%%%%%%%%%%%%%%%%%%%%%%%%%%%%%%%%%%%%%%%%%%%%%%%%%%%%%%%
\begin{abstract}
We investigate global locations of Schmidt number witnesses which
are outside of the convex set of all bipartite states. Their
locations are classified by interiors of faces of the convex set of all states, by
considering the line segments from them to the maximally mixed
state. In this way, a nonpositive Hermitian matrix of trace 1 is
located outside of one and only one face. Faces of the convex set of
all states are classified by subspaces, which are range spaces of
states belonging to specific faces. For a given subspace, we show
that there exist Schmidt number $k+1$ witnesses outside of the face
arising from this subspace if and only if every vector in the
orthogonal complement of the subspace has Schmidt rank greater than
$k$. Once we have Schmidt number
$k+1$ witnesses outside of a face, we also have Schmidt number
$2,3,\dots, k$ witnesses outside of the face.
\end{abstract}

\maketitle

%%%%%%%%%%%%%%%%%%%%%%%%%%%%%%%%%%%%%%%%%%%%%%%%%%%%%%%%%%%%%%%%%%%%%
%%%%%%%%%%%%%%%%%%%%%%%%%%%%%%%%%%%%%%%%%%%%%%%%%%%%%%%%%%%%%%%%%%%%%%%%%%%%%%%%%%%%%%%%%
%%%%%%%%%%%%%%%%%%%%%%%%%%%%%%%%%%%%%%%%%%%%%%%%%%%%%%%%%%%%%%%%%%%%%%%%%%%%%%%%%%%%%%%%%
%%%%%%%%%%%%%%%%%%%%%%%%%%%%%%%%%%%%%%%%%%%%%%%%%%%%%%%%%%%%%%%%%%%%%%%%%%%%%%%%%%%%%%%%%
%%%%%%%%%%%%%%%%%%%%%%%%%%%%%%%%%%%%%%%%%%%%%%%%%%%%%%%%%%%%%%%%%%%%%%%%%%%%%%%%%%%%%%%%%
%%%%%%%%%%%%%%%%%%%%%%%%%%%%%%%%%%%%%%%%%%%%%%%%%%%%%%%%%%%%%%%%%%%%%%%%%%%%%%%%%%%%%%%%%
\section{Introduction}
%%%%%%%%%%%%%%%%%%%%%%%%%%%%%%%%%%%%%%%%%%%%%%%%%%%%%%%%%%%%%%%%%%%%%
The notion of entanglement \cite{Werner-1989} for mixed states is
now playing the key role in the current quantum communication and quantum information theory \cite{{beng_zyc},{guhne},{horo-survey}}.
Schmidt numbers \cite{terhal-schmidt} of bipartite states measure
the degree of entanglement; the higher Schmidt number a state has, the more entangled it is;
a state has Schmidt number 1 if and only if it is not entangled at all.
It is very difficult in general to determine the Schmidt number of a given state \cite{gurvits}, and we
need Schmidt number $k$ witnesses \cite{SBL_2001} for this purpose.
During the last couple of years, much attention has been paid to understanding Schmidt number witnesses
%and understand the structures of them
\cite{{WCKMB_2023},{ZDAG_2024},{Krebs_Gach_2024},{shi_2024},{BBRTB_2024},{DCT_2025},
{TavMor_2024},{LHF_2025},{LLFH_2025},{MMDKG_2025}}, together with $k$-positive maps
\cite{{park_youn_2024},{ADMPR_pre},{mlynik_osaka_marcin_2025}} whose Choi matrices \cite{choi75-10}
play roles of Schmidt number $k+1$ witnesses.

In this paper, we investigate the locations of Schmidt number witnesses, which are nonpositive
bipartite Hermitian matrices. For this purpose, we consider the hyperplane $H$ of all Hermitian matrices in $M_m\ot M_n$
with trace 1, on which the convex set $\mathcal D$ of all $m\ot n$ bipartite states is located.
Then Schmidt number witnesses are located outside of $\mathcal D$ on $H$. In order to specify the locations
of Schmidt number witnesses, we partition the outside of $\mathcal D$ by the interiors of faces of $\mathcal D$.
By the {\sl interior} of a convex set, we mean the interior with respect to the relative topology
induced by the hyperplane generated by the convex set. Then every convex set has a nonempty interior.
In particular, every face has its own nonempty interior.
For example, the interior of an extreme point is itself.
A point in the interior is called an {\sl interior point}, and a point which is not an interior point
is called a {\sl boundary point}. A boundary point of the convex set $\mathcal D$ will be called a {\sl boundary state}.

For a given nonpositive Hermitian matrix $X$ with trace 1 and a nontrivial face $F$ of $\mathcal D$,
we say that  $X$ is located {\sl outside of $F$}
when the line segment from $X$ to the maximally mixed state $\varrho_*:=\frac 1{mn}I_{mn}$
meets the interior of $F$. Since $\varrho_*$ is an interior point of $\mathcal D$ \cite{Gurvits_Barnum} and
the boundary of a convex set is completely partitioned into the interiors of nontrivial faces \cite{rock},
we see that $X$ is located outside of one and only one face of $\mathcal D$.
Such a partition of the outside of a convex set is always possible by choosing an interior point.
For example, the outside of a triangle is partitioned by six regions, depending on the choice of interior point;
three of them are two-dimensional regions
which are outside of edges, and other three of them are one-dimensional regions outside of extreme points.

We recall that a subspace $E$ of $\mathbb C^m\ot\mathbb C^n$ gives rise to the face $F_E$ of the convex set $\mathcal D$,
which consists of all states whose ranges are contained in $E$, and every face arises in this way.
The interior of $F_E$ consists of states whose ranges are precisely $E$.
We show that there exist Schmidt number $k+1$ witnesses outside of the face $F_E$ if and only if
every vector in the orthogonal complement $E^\perp$ has Schmidt rank greater than $k$; in other words,
$E^\perp$ is  {\sl $k$-entangled} \cite{lovitz_jognson}.
A $1$-entangled subspace is just a completely entangled subspace
in the sense of \cite{Parthasarathy_2004}, which has no product vector.

Note that every $k$-entangled subspace is $\ell$-entangled whenever $\ell \le k$.
We say that $E$ is an {\sl exactly $k$-entangled subspace} when $k$ is the greatest number
such that $E$ is $k$-entangled.
Then completely entangled subspaces are classified by exactly $k$-entangled subspaces
with $k=1,2,\dots, m\meet n-1$, where $m\meet n$ denotes the minimum of $\{m,n\}$. We see that $E^\perp$ is
exactly $k$-entangled if and only if $k$ is the greatest number such that there are Schmidt number $k+1$ witnesses outside of $F_E$.
When this is the case, there exist Schmidt number $\ell$ witnesses outside of $F_E$ for every $\ell=2,3,\dots, k$.

We recall that a point $x$ of a convex set $C$ is an interior point if and only if
every line segment from a point in $C$ to $x$ can be extended within $C$. See \cite{rock}.
When we fix an interior point $x_0$ of a convex set $C$, we note \cite{kye-canad}
that a point $x\in C$ is an interior point of $C$ if and only if
the line segment from $x_0$ to $x$ can be extended within $C$.
In this context, we consider the one parameter family,
$$
X_\lambda=(1-\lambda)\varrho_*+\lambda\varrho,\qquad -\infty<\lambda<+\infty,
$$
of Hermitian matrices for a given bipartite state $\varrho$, which will be the main tool for our investigations.
We will use the notation $X^\varrho_\lambda$
when we emphasize the role of $\varrho$.
If $\varrho$ is the maximally entangled state, then $X_\lambda$'s
and their partial transposes give rise to the isotropic states
\cite{terhal-schmidt} and the Werner states \cite{Werner-1989},
respectively. If we take the linear map between matrices whose Choi
matrix is given by $X^\varrho_\lambda$ with the maximally entangled
state $\varrho$, then we get the one parameter family of
Tomiyama maps \cite{tom_85} which distinguish $k$-positivity for
different $k$'s. This one parameter family includes Choi's
example \cite{choi72} of an $(n-1)$-positive map, which is the first example
in the literature distinguishing $k$-positivity and complete positivity.

We recall that $X$ is a Schmidt number $k+1$ witness when $X$ is
$k$-blockpositive but not $k+1$-blockpositive. We first get the
formulas to identify the boundary of the convex set $\blockpos_k$
consisting of all $k$-blockpositive matrices of trace 1 in terms
of $X_\lambda$. Then we proceed to interpret geometric meanings of
these formulas to get the results mentioned above. We will
illustrate our results on the two-dimensional planes.

Parts of this paper were presented by the second author at \lq\lq Mathematical Structures
in Quantum Mechanics\rq\rq\ which was held at Gdansk, Poland, in March 2025.
He is grateful to the audience for stimulating discussions and organizers Adam Rutkowski and Marcin Marciniak for travel support.

%%%%%%%%%%%%%%%%%%%%%%%%%%%%%%%%%%%%%%%%%%%%%%%%%%%%%%%%%%%%%%%%%%%%%
%%%%%%%%%%%%%%%%%%%%%%%%%%%%%%%%%%%%%%%%%%%%%%%%%%%%%%%%%%%%%%%%%%%%%%%%%%%%%%%%%%%%%%%%%
%%%%%%%%%%%%%%%%%%%%%%%%%%%%%%%%%%%%%%%%%%%%%%%%%%%%%%%%%%%%%%%%%%%%%%%%%%%%%%%%%%%%%%%%%
%%%%%%%%%%%%%%%%%%%%%%%%%%%%%%%%%%%%%%%%%%%%%%%%%%%%%%%%%%%%%%%%%%%%%%%%%%%%%%%%%%%%%%%%%
%%%%%%%%%%%%%%%%%%%%%%%%%%%%%%%%%%%%%%%%%%%%%%%%%%%%%%%%%%%%%%%%%%%%%%%%%%%%%%%%%%%%%%%%%
%%%%%%%%%%%%%%%%%%%%%%%%%%%%%%%%%%%%%%%%%%%%%%%%%%%%%%%%%%%%%%%%%%%%%%%%%%%%%%%%%%%%%%%%%
\section{Schmidt number witnesses}
%%%%%%%%%%%%%%%%%%%%%%%%%%%%%%%%%%%%%%%%%%%%%%%%%%%%%%%%%%%%%%%%%%%%%%%%%%%%%%%%%%%%%%%%%
We collect all bipartite pure states whose range vectors have
Schmidt rank $\le k$, and consider the convex hull $\mathcal S_k$ of
them \cite{eom-kye}. A  state is said to be of {\sl Schmidt number $k$}
\cite {terhal-schmidt} if it belongs to $\mathcal
S_k\setminus\mathcal S_{k-1}$. A Hermitian matrix $W\in M_m\ot M_n$
is called {\sl $k$-blockpositive} \cite{{jam_72},{ssz}} if
$\lan\xi|W|\xi\ran\ge 0$ for every $|\xi\ran$ with $\sr|\xi\ran\le
k$, where $\sr|\xi\ran$ denotes the Schmidt rank of $|\xi\ran\in
\mathbb C^m\ot\mathbb C^n$.
Then we have the chain
\begin{equation}\label{chain}
\mathcal S_1\subset\cdots\subset \mathcal
S_{m\meet n} =\blockpos_{m\meet
n}\subset\cdots\subset\blockpos_1
\end{equation}
of inclusions of compact convex sets of Hermitian matrices in
$M_m\ot M_n$, where $\mathcal S_{m\meet n}=\blockpos_{m\meet n}$
coincides with the convex set $\mathcal D$ of all bipartite states. Note that a
state $\varrho$ belongs to $\mathcal S_k$ if and only if $\lan
\varrho|W\ran\ge 0$ for every $W\in\blockpos_k$. Therefore, we see that $\varrho$ has
Schmidt number $\ge k$ if and only if $\varrho\notin \mathcal
S_{k-1}$ if and only if there exists $W\in\blockpos_{k-1}$ such that
$\lan\varrho| W\ran<0$. If $\varrho$ has Schmidt number $k$ then
such a $W$ belongs to $\blockpos_{k-1}\setminus\blockpos_{k}$. A
Hermitian matrix $W\in\blockpos_{k-1}\setminus\blockpos_{k}$ is
called a {\sl Schmidt number $k$ witness} \cite{SBL_2001}.

We recall that a linear map is $k$-positive \cite{stine} if and only if its
Choi matrix \cite{choi75-10} is $k$-blockpositive, and so we may also detect
Schmidt numbers by $k$-positive maps, using the usual duality between mapping spaces and
tensor products. When $k=1$, this gives rise to the Horodecki's separability criterion \cite{horo-1}
using positive maps. On the other hand, we need states of Schmidt number $\le k$, in order to
determine $k$-positivity of a linear map between matrices \cite{eom-kye}.
For expository articles on the related topics, see
\cite{{bruss_sur},{horo-survey},{kye_comp_tensor},{kye_lec_note}}
for examples. We also refer to \cite{{Shirokov_2013},{han_kye_str}} for infinite-dimensional analogs
of the relations between Schmidt numbers and $k$-positivity.

We see that $X_\lambda$ is $k$-blockpositive if and only if the inequality
$(1-\lambda) \frac 1{mn}\ge -\lambda \lan\xi|\varrho|\xi\ran$
holds for every unit vector $|\xi\ran\in\mathbb C^m\ot\mathbb C^n$
with $\sr|\xi\ran\le k$. Therefore, it is natural to consider the
supreme of $\lan\xi|\varrho|\xi\ran$ under the condition $\sr|\xi\ran\le
k$, as in \cite[Proposition 2]{SBL_2001}. The authors of \cite{jonston_kribs} considered the supreme of
$|\lan\xi|X|\eta\ran|$ under the same rank conditions of unit
vectors for  arbitrary $X\in M_m\ot M_n$, and called that the {\sl $k$-th operator norm}
$\|X\|_{S(k)}$. See also \cite{{GPMSZ},{GM_2015}}. For positive $\varrho$,
it was also shown in \cite{jonston_kribs} that
\begin{equation}\label{sk_def}
\|\varrho\|_{S(k)} = \sup\{\lan\xi|\varrho|\xi\ran:
\|\xi\|=1,\ \sr|\xi\ran\le k\}.
\end{equation}
When $k=m\meet n$, we see that
$\|\varrho\|_{S(m\meet n)}=\|\varrho\|$
is the greatest eigenvalue of $\varrho$.

In order to find lower bounds of $\lambda$ satisfying $X_\lambda\in\blockpos_k$,
we consider the case $\lambda<0$. Then we see that
$X_\lambda$ is $k$-blockpositive if and only if
$-1 \le (mn ||\varrho||_{S(k)} -1)\lambda$. We have $mn ||\varrho||_{S(k)}-1 > 0$. Otherwise,
$X_\lambda$ is $k$-blockpositive for all $\lambda<0$, which contradicts that $\mathcal{BP}_k$ is compact.
Therefore, we conclude
\begin{equation}\label{beta-}
\beta^-_k:={-1 \over mn ||\varrho||_{S(k)} -1}\le\lambda
\ \Longleftrightarrow\
X_\lambda\in\blockpos_k.
\end{equation}

In order to find upper bounds, we also consider the number
\begin{equation}\label{def_|.|}
|\varrho|_{S(k)} := \inf \{\lan \xi | \varrho | \xi \ran : ||\xi||=1, {\rm SR}(\xi) \le k \}.
\end{equation}
By  $\lan \xi |~ ||\varrho|| I_{mn} - \varrho |\xi \ran = ||\varrho|| - \lan \xi | \varrho | \xi \ran$,
we have
\begin{equation}\label{id_|.|}
|\varrho|_{S(k)}
= || \varrho || - \left\|~ ||\varrho|| I_{mn} - \varrho\right\|_{S(k)}.
\end{equation}
Note that $|\varrho|_{S(m\meet n)}$ is the lowest eigenvalue of $\varrho$.
When $\lambda>0$, we see that $X_\lambda$ is $k$-blockpositive if and only if
$0 \le  {1-\lambda \over mn} + \lambda \lan \xi | \varrho | \xi \ran$
for all unit vectors $|\xi\ran$ with Schmidt rank $\le k$ if and only if
$\left(1 - mn |\varrho|_{S(k)}\right) \lambda \le 1$. We have
$1 - mn |\varrho|_{S(k)} > 0$ as before, and conclude
\begin{equation}\label{beta+}
X_\lambda\in\blockpos_k\
\Longleftrightarrow\
\lambda\le {-1 \over mn |\varrho|_{S(k)}-1} =: \beta^+_k.
\end{equation}

When we consider the witness of the form $I_{mn}-\alpha\varrho$, or $\alpha I_{mn}-\varrho$ as in \cite{SBL_2001}, we have
$\alpha I_{mn}-\varrho\in\blockpos_k$ if and only if $\|\varrho\|_{S(k)}\le \alpha$, and so we have
\begin{equation}\label{form_other}
\|\varrho\|_{S(k)}=\min\{\alpha: \alpha I_{mn}-\varrho\in\blockpos_k\}.
\end{equation}
A number $\alpha$ satisfying $\alpha I_{mn}-\varrho\in\blockpos_k$ has been found in \cite{ZDAG_2024} for $k=1,2,3$,
in terms of operator Schmidt decomposition \cite{NDDGMOBH_2003}. Using the identity (\ref{form_other}),
the number $\|\varrho\|_{S(k)}$ can be computed by  \cite[Theorem 3.2]{mlynik_osaka_marcin_2025},
with the correspondence between $k$-blockpositive matrices and $k$-positive maps.

We proceed to get a formula for $\|\varrho\|_{S(k)}$ in terms of Schmidt coefficients of the range vectors of $\varrho^{1/2}$,
which is fit for our purpose.
For a vector $|\xi\ran\in\mathbb C^m\ot\mathbb C^n$, we consider the Schmidt coefficients $s_1\ge \cdots\ge s_{m\meet n}$
which are nothing but singular values of the $m\times n$ matrix corresponding to the vector $|\xi\ran\in\mathbb C^m\ot\mathbb C^n$.
We put
$$
\tau_k|\xi\ran=s_1^2+s_2^2+\cdots+s_k^2.
$$
Then $|\xi\ran\mapsto \tau_k|\xi\ran$ defines a continuous function on the space $\mathbb C^m\ot\mathbb C^n$.
When we restrict the domain to the unit sphere $\mathbb S$, it takes the minimum $\frac k{m\meet n}$
when Schmidt coefficients are evenly distributed,
and takes the maximum $1$ when $\SR|\xi\ran\le k$.

In order to get a formula to compute $\|\varrho\|_{S(k)}$, we first recall
that $\lan\eta|\varrho|\eta\ran=\|\varrho^{1/2}|\eta\ran\|^2$, which is the supreme of
$|\lan\xi|\varrho^{1/2}|\eta\ran|^2=|\lan\varrho^{1/2}\xi|\eta\ran|^2$ with $|\xi\ran\in\mathbb S$. Therefore, we see that
$\|\varrho\|_{S(k)}$ is the supreme of $|\lan\xi|\eta\ran|^2$ with
$|\xi\ran\in\varrho^{1/2}(\mathbb S)$, $|\eta\ran\in\mathbb S$ and $\SR|\eta\ran\le k$.
It was shown in \cite[Theorem 3.3]{jonston_kribs} that $\tau_k|\xi\ran$ is the supreme of $|\lan\xi|\eta\ran|^2$ with
$|\eta\ran\in\mathbb S$ and $\SR|\eta\ran\le k$ (see the Appendix for an elementary proof). Therefore, we have
\begin{equation}\label{formu}
\|\varrho\|_{S(k)}=\sup\{ \tau_k|\xi\ran : |\xi\ran \in \varrho^{1/2}(\mathbb S)\}.
\end{equation}
%for a state $\varrho$.
When $\varrho_E$ is a {\sl projection state} onto a subspace $E$, that is, $\varrho_E=\frac 1{\dim E}P_E$ with the projection
$P_E$ onto $E$, we see that
$\|\varrho_E\|_{S(k)}$ is given by the supreme of $\frac 1{\dim E}\tau_k|\xi\ran$ with unit vectors $|\xi\ran\in E$. This recovers
formulas in \cite{{PPHH},{jonston_kribs_2}}.
One may recover a typical form of \cite[Theorem 3.2]{mlynik_osaka_marcin_2025}
from (\ref{formu}) by the spectral decomposition of $\varrho$ and the Choi correspondence.

Formula (\ref{formu}) is useful to compute $\|\varrho\|_{S(k)}$.
When $m=n$ and $\varrho$ is the maximally entangled state for example, we have $\|\varrho\|_{S(k)}=\frac kn$, and so $\beta^-_k=\frac {-1}{nk-1}$,
which recovers Tomiyama's result \cite{tom_85}, by the correspondence between $k$-blockpositive matrices and $k$-positive maps.
We also consider the projection state $\varrho=\frac 2{n(n-1)}\sum_{i>j}|\xi_{ij}\ran\lan\xi_{ij}|$
onto the antisymmetric space, with $|\xi_{ij}\ran=\frac 1{\sqrt 2}(|i\ran|j\ran-|j\ran|i\ran)$.
Then we have $\|\varrho\|_{S(1)}=\frac 1{n(n-1)}$ and $\|\varrho\|_{S(k)}=\frac 2{n(n-1)}$ for $k=2,\dots,n$,
and so $\beta^-_1=-(n-1)$ and $\beta^-_k=-\frac{n-1}{n+1}$ for $k=2,\dots,n$.
For $\lambda={\beta^-_1}$, we note that $X^\varrho_\lambda$ is the Choi matrix of the transpose map up to a scalar multiplication,
and so we also recover the result in \cite{tom_85} for $k$-copositivity by reparametrization.
See {\sc Fig.\ 1.2} in \cite[Section 1.7]{kye_lec_note}.

%%%%%%%%%%%%%%%%%%%%%%%%%%%%%%%%%%%%%%%%%%%%%%%%%%%%%%%%%%%%%%%%%%%%%
%%%%%%%%%%%%%%%%%%%%%%%%%%%%%%%%%%%%%%%%%%%%%%%%%%%%%%%%%%%%%%%%%%%%%%%%%%%%%%%%%%%%%%%%%
%%%%%%%%%%%%%%%%%%%%%%%%%%%%%%%%%%%%%%%%%%%%%%%%%%%%%%%%%%%%%%%%%%%%%%%%%%%%%%%%%%%%%%%%%
%%%%%%%%%%%%%%%%%%%%%%%%%%%%%%%%%%%%%%%%%%%%%%%%%%%%%%%%%%%%%%%%%%%%%%%%%%%%%%%%%%%%%%%%%
%%%%%%%%%%%%%%%%%%%%%%%%%%%%%%%%%%%%%%%%%%%%%%%%%%%%%%%%%%%%%%%%%%%%%%%%%%%%%%%%%%%%%%%%%
%%%%%%%%%%%%%%%%%%%%%%%%%%%%%%%%%%%%%%%%%%%%%%%%%%%%%%%%%%%%%%%%%%%%%%%%%%%%%%%%%%%%%%%%%
\section{Geometric Interpretations}
%%%%%%%%%%%%%%%%%%%%%%%%%%%%%%%%%%%%%%%%%%%%%%%%%%%%%%%%%%%%%%%%%%%%%%%%%%%%%%%%%
We have seen that $X_\lambda$ is $k$-blockpositive if and only if $\beta^-_k\le\lambda\le\beta^+_k$.
We put $\delta^-:=\beta^-_{m\meet n}$ and $\delta^+:=\beta^+_{m\meet n}$, then we see that $X_\lambda$ is a state if and only if
$\delta^-\le\lambda\le\delta^+$.
Since the function $t\mapsto \frac {-1}{mnt-1}$ is increasing on the intervals $(-\infty,\frac 1{mn})$
and $(\frac 1{mn},+\infty)$, we have
$$
\begin{aligned}
&\beta^-_1\le\beta^-_2\le\cdots\le \beta^-_{m\meet n-1}\le\delta^-<0,\\
&1\le\delta^+\le\beta^+_{m\meet n-1}\le\cdots\le\beta^+_2\le\beta^+_1.
\end{aligned}
$$

We first consider the question when the strict inequalities $\beta^-_k<\delta^-$ and/or $\delta^+<\beta^+_k$
hold, in order to find possible locations of Schmidt number witnesses.
Since $\varrho_*$ is an interior point of $\blockpos_k$, we see that
the boundary state $X_{\delta^-}$ is an interior point of $\blockpos_k$ if and only if
$\beta^-_k<\delta^-$ holds.
We also note that $\beta^-_k=\delta^-$ if and only if $||\varrho||_{S(k)}=||\varrho||$ if and only if
there exists a unit vector $|\xi\ran \in \mathbb C^m \otimes \mathbb C^n$ such that $\SR|\xi\ran\le k$ and $\lan\xi|\varrho|\xi\ran=||\varrho||$
if and only if there exists a unit vector $|\xi\ran$ with ${\rm SR}|\xi\ran\le k$ in the eigenspace corresponding
to the greatest eigenvalue. Therefore, we have (i) of the following;

\begin{theorem}\label{boundary-}
For a state $\varrho$, we have the following:
\begin{enumerate}
\item[{\rm (i)}]
The boundary state $X^\varrho_{\delta^-}$ is an interior point of $\mathcal{BP}_k$ if and only if
$\|\varrho\|_{S(k)}<\|\varrho\|$ if and only if
the eigenspace corresponding to the greatest eigenvalue of $\varrho$ is $k$-entangled.
\item[{\rm (ii)}]
The boundary state $X^\varrho_{\delta^+}$  is an interior point of $\mathcal{BP}_k$ if and only if
$|\varrho|_{S(k)}>|\varrho|_{S(m\meet n)}$ if and only if
the eigenspace corresponding to the lowest eigenvalue of $\varrho$ is $k$-entangled.
\end{enumerate}
\end{theorem}

The statement (ii) can be seen by exactly the same argument using $|\varrho|_{S(k)}$ instead of $\|\varrho\|_{S(k)}$.
We note that there exist $k$-blockpositive matrices outside of the face $F_E$
if and only if the interior of $F_E$ is contained in the interior of $\blockpos_k$ if and only if
there is an interior point of $F_E$ which is an interior point of $\blockpos_k$.
The projection state $\varrho_E$ is an interior point of the face $F_E$, and so we apply Theorem \ref{boundary-} (ii)
to have the following;

\begin{theorem}\label{boundary-2}
There exists a $k$-blockpositive matrix outside of a face $F_E$ if and only if
$E^\perp$ is $k$-entangled.
\end{theorem}

Now, we consider a face $F_E$ and its \lq\lq opposite\rq\rq\ face
$F_{E^\perp}$. We also consider the  convex hull $\Delta$ of
these two faces, and the hyperplane $H$ consisting of Hermitian
matrices generated by two faces.
Then it is easy to see that a Hermitian matrix on the hyperplane $H$ is a state if and only if it belongs to $\Delta$.
Considering the projection state $\varrho_E$, we have
$\lan \varrho-\varrho_E|\varrho_*-\varrho_E\ran=0$ for every $\varrho\in F_E$, and so we see that
the face $F_E$ is perpendicular to the one parameter family $\{X_\lambda^{\varrho_E}\}$.
The opposite face $F_{E^\perp}$ is also perpendicular to the one parameter family $\{X_\lambda^{\varrho_E}\}$
at the intersection point $\varrho_{E^\perp}$.
See {\sc FIG.\ 1}.

\begin{figure}
\begin{center}
\setlength{\unitlength}{2 truecm}
\begin{picture}(1.5,1.5)
\thinlines
\drawline(0.5,1.5)(0,0.5)
\dottedline{0.05}(0,0.5)(1.4,0.3)
%\drawline(0,0.5)(0.2,0)
\drawline(0.2,0)(1.4,0.3)
\drawline(0.5,1.5)(0.2,0)
\dottedline{0.05}(0,0.5)(1.4,0.3)
\linethickness{0.4mm}\dottedline{0.005}(1.4,0.3)(0.5,1.5)
\linethickness{0.4mm}\dottedline{0.005}(0,0.5)(0.2,0)
\put(0.95,0.9){\circle*{0.07}}
\put(0.1,0.25){\circle*{0.07}}
\put(1.46,1.29){\circle*{0.07}}
\put(0.44,0.51){\circle*{0.07}}
\put(-0.3675,-0.1075){\circle*{0.07}}
\linethickness{0.4mm}\dottedline{0.07}(1.63,1.42)(-0.495,-0.202)
\put(1.05,0.85){$\varrho_E=X_1$}
\put(0.51,0.48){$\varrho_*=X_0$}
\put(-0.92,0.22){$\varrho_{E^\perp}=X_{\delta^-}$}
\put(1.52,1.18){$X_{\beta^+_k}$}
\put(-0.3,-0.18){$X_{\beta^-_k}$}
\end{picture}
\vskip 1pc
\end{center}
\caption{
Thick lines in this figure represent the face $F_E$
and its opposite face $F_{E^\perp}$. The tetrahedron represents
the set of all states among Hermitian matrices located on the  hyperplane generated by these two faces.
The dotted line represents the one parameter family $X_\lambda^{\varrho_E}$,
with five dots for $\lambda=\beta^+_k$, $\delta^+=1$, $0$, $\delta^-$ and $\beta^-_k$ from the top to the bottom.
}
\end{figure}

Theorem \ref{boundary-2} tells us that there exist Schmidt number $k+1$ witnesses outside of
the face $F_E$ only  when $E^\perp$ is $k$-entangled.
In order to confirm the existence of Schmidt number $k+1$ witnesses outside of $F_E$ and/or $F_{E^\perp}$, we have to consider
the strict inequalities $\beta^-_{k}<\beta^-_{k+1}$ and/or $\beta^+_{k+1}<\beta^+_{k}$.
The identity (\ref{formu}) is useful for this purpose.
Suppose that the range space of $\varrho$ is $k$-entangled and $\ell\le k$.
Take $|\xi\ran \in \varrho^{1/2}(\mathbb S)$ such that $\|\varrho\|_{S(\ell)}=\tau_\ell|\xi\ran$.
Then we have $\SR|\xi\ran >k\ge \ell$ and so $\tau_{\ell}|\xi\ran<\tau_{\ell+1}|\xi\ran$,
which implies $\|\varrho\|_{S(\ell)} < \|\varrho\|_{S(\ell+1)}$.
This proves (i) of the following;

\begin{theorem}\label{strct-ineq}
For a state $\varrho$ in $M_m \otimes M_n$, we have the following:
\begin{enumerate}
\item[{\rm (i)}]
If the range space is $k$-entangled, then we have
$\beta^-_1 < \cdots < \beta^-_{k} <  \beta^-_{k+1}$.
\item[{\rm (ii)}]
If the orthogonal complement of the eigenspace corresponding to the greatest eigenvalue is $k$-entangled,
then we have
$\beta^+_{k+1}<\beta^+_{k}<\cdots<\beta^+_1$.
\end{enumerate}
\end{theorem}

For the statement (ii), we put $\sigma:=\|\varrho\|I_{mn}-\varrho$. Then the assumption of (ii)
implies that the range of $\sigma$ is $k$-entangled, and so $\|\sigma\|_{S(\ell)}<\|\sigma\|_{S(\ell+1)}$ whenever $\ell\le k$.
By the identity (\ref{id_|.|}), we have $|\varrho|_{S(\ell+1)} < |\varrho|_{S(\ell)}$.

For a given face $F_E$ of the convex set $\mathcal D$, suppose that
there is a Schmidt number $k+1$ witness outside of $F_E$ then $E^\perp$ is $k$-entangled by
Theorem \ref{boundary-2}. In this case, there exist Schmidt number $\ell$ witnesses for $\ell=2,\dots, k$,
by applying Theorem \ref{strct-ineq} (ii) for the projection state $\varrho_E$.
Therefore, we have the following:

\begin{theorem}\label{conc}
For a subspace $E$ of $\mathbb C^m\ot \mathbb C^n$, the orthogonal complement $E^\perp$
is exactly $k$-entangled  if and only if
$k$ is the greatest number such that there exist Schmidt number $k+1$ witnesses outside of $F_E$.
When this is the case, there exist Schmidt number $\ell$ witnesses outside of $F_E$ for $\ell=2,3,\dots, k$.
\end{theorem}

It is known \cite{CMW08}
that if a subspace $E$ of $\mathbb C^m\ot\mathbb C^n$ satisfies $\dim E>
(m-k)(n-k)$ then there exists a vector $|\xi\ran$ in $E$ with
$\sr|\xi\ran\le k$. Therefore,
there exist Schmidt number $k+1$ witnesses outside of the face $F_E$ only when
$\dim E\ge k(m+n-k)$.
In the $n\ot n$ system, there exist Schmidt number $n$ witnesses
outside of the face $F_E$ when and only when $E^\perp$ is of one dimension
with a vector of maximal Schmidt rank, as we see in Choi's 1972 example \cite{choi72}
of $(n-1)$ positive map which is not completely positive. On the other hand, there exists no witness
outside of a pure state which is an extreme point of $\mathcal D$. In fact,
it is known that every pure state is an exposed extreme point of the much bigger convex set $\blockpos_1$
\cite{{marcin_exp},{kye_exp}}.

%%%%%%%%%%%%%%%%%%%%%%%%%%%%%%%%%%%%%%%%%%%%%%%%%%%%%%%%%%%%%%%%%%%%%
%%%%%%%%%%%%%%%%%%%%%%%%%%%%%%%%%%%%%%%%%%%%%%%%%%%%%%%%%%%%%%%%%%%%%%%%%%%%%%%%%%%%%%%%%
%%%%%%%%%%%%%%%%%%%%%%%%%%%%%%%%%%%%%%%%%%%%%%%%%%%%%%%%%%%%%%%%%%%%%%%%%%%%%%%%%%%%%%%%%
%%%%%%%%%%%%%%%%%%%%%%%%%%%%%%%%%%%%%%%%%%%%%%%%%%%%%%%%%%%%%%%%%%%%%%%%%%%%%%%%%%%%%%%%%
%%%%%%%%%%%%%%%%%%%%%%%%%%%%%%%%%%%%%%%%%%%%%%%%%%%%%%%%%%%%%%%%%%%%%%%%%%%%%%%%%%%%%%%%%
%%%%%%%%%%%%%%%%%%%%%%%%%%%%%%%%%%%%%%%%%%%%%%%%%%%%%%%%%%%%%%%%%%%%%%%%%%%%%%%%%%%%%%%%%
\section{Illustration}
%%%%%%%%%%%%%%%%%%%%%%%%%%%%%%%%%%%%%%%%%%%%%%%%%%%%%%%%%%%%%%%%%%%%%%%%%%%%%
Take $|\xi_1\ran=|01\ran$, $|\xi_2\ran = {1\over \sqrt{2}}(|01\ran+|10\ran)$ and
$|\xi_3\ran = {1\over \sqrt{3}}(|00\ran+|11\ran+|22\ran)$ in $\mathbb C^3\ot\mathbb C^3$,
together with the corresponding projection states $\varrho_1$, $\varrho_2$ and $\varrho_3$.
We also consider the two-dimensional plane $H_i$ through $\varrho_3$, $\varrho_i$ and $\varrho_*$ for $i=1,2$.
In order to illustrate our results, we will determine the regions for Schmidt number $k$ witnesses on the plane $H_i$ for $i=1,2$.
We note that the line segment $\overline{\varrho_3 \varrho_i}$ is contained in the face of $\mathcal D$
determined by the two dimensional subspace spanned by $|\xi_3\ran$ and $|\xi_i\ran$.
The opposite face is determined by its orthogonal complement with the associated
projection state $\sigma_i:={1 \over 7}(9\varrho_* - \varrho_3 - \varrho_i)$. Compare with {\sc FIG.\ 1}.
Therefore, the region for states on the plane $H_i$ is surrounded by the triangle
with vertices $\varrho_3$, $\varrho_i$ and $\sigma_i$ for $i=1,2$.  See {\sc FIG.\ 2}.

\begin{figure}
\begin{center}
\setlength{\unitlength}{2.5 truecm}
\begin{picture}(2.5,1.5)
\thinlines
\put(0,1.41421){\circle*{0.03}}%A
\put(0,0){\circle*{0.03}}%B
\put(0.80178,0.70710){\circle*{0.03}}%C
\put(0.62360,0.70710){\circle*{0.03}}%O
\put(0.70156,0.61871){\circle*{0.03}}%E
\put(0.70156,0.79549){\circle*{0.03}}%F
\put(0.93541,0.35355){\circle*{0.03}}%I
\put(0.74833,0.56568){\circle*{0.03}}%J
\dottedline{0.005}(0,1.41421356237309)(0,0)
\dottedline{0.005}(0,0)(0.801783725737273,0.707106781186547)
\dottedline{0.005}(0.801783725737273,0.707106781186547)(0,1.41421356237309)
%\dottedline{0.05}(0,0)(0.801783725737273,0.90913729009699)
%\dottedline{0.05}(0,0.707106781186547)(1.12249721603218,0.707106781186547)
\dottedline{0.05}(0,1.41421356237309)(0.935414346693485,0.353553390593274)
%\dottedline{0.005}(0,1.41421356237309)(0.801783725737273,0.90913729009699)
%\dottedline{0.005}(0.801783725737273,0.90913729009699)(0.935414346693485,0.353553390593274)
\dottedline{0.005}(0.935414346693485,0.353553390593274)(0,0)
\dottedline{0.005}(0,0)(0.748331477354788,0.565685424949238)

\put(-0.15,1,4){$\varrho_3$}
\put(-0.15,0){$\varrho_1$}
\put(0.47,0.67){$\varrho_*$}
\put(0.81,0.73){$\sigma_1$}

\put(1.5,1.41421){\circle*{0.03}}%A
\put(1.5,0){\circle*{0.03}}%B
\put(2.30178,0.70710){\circle*{0.03}}%C
\put(2.12360,0.70710){\circle*{0.03}}%O
%\put(1.5,0.70710){\circle*{0.03}}%D
\put(2.20156,0.61871){\circle*{0.03}}%E
\put(2.20156,0.79549){\circle*{0.03}}%F
\put(2.30178,0.90913){\circle*{0.03}}%G
\put(2.43541,0.35355){\circle*{0.03}}%I
\put(2.24833,0.56568){\circle*{0.03}}%J

\dottedline{0.005}(1.5,1.41421356237309)(1.5,0)
\dottedline{0.005}(1.5,0)(2.30178372573727,0.707106781186547)
\dottedline{0.005}(2.30178372573727,0.707106781186547)(1.5,1.41421356237309)
%\dottedline{0.05}(1.5,0)(2.30178372573727,0.90913729009699)
%\dottedline{0.05}(1.5,0.707106781186547)(2.62249721603218,0.707106781186547)
\dottedline{0.05}(1.5,1.41421356237309)(2.43541434669349,0.353553390593274)

\dottedline{0.005}(1.5,1.41421356237309)(2.30178372573727,0.90913729009699)
\dottedline{0.005}(2.30178372573727,0.90913729009699)(2.43541434669349,0.353553390593274)
\dottedline{0.005}(2.43541434669349,0.353553390593274)(1.5,0)
\dottedline{0.005}(1.5,0)(2.24833147735479,0.565685424949238)

\put(1.34,1,4){$\varrho_3$}
\put(1.34,0){$\varrho_2$}
\put(1.98,0.67){$\varrho_*$}
\put(2.31,0.73){$\sigma_2$}

%3x3_sn=1%BP1
\put(0.80178,0.70710){\circle*{0.001}}
\put(0.80648,0.70295){\circle*{0.001}}
\put(0.81144,0.69858){\circle*{0.001}}
\put(0.81667,0.69397){\circle*{0.001}}
\put(0.82221,0.68909){\circle*{0.001}}
\put(0.82807,0.68392){\circle*{0.001}}
\put(0.83428,0.67844){\circle*{0.001}}
\put(0.84089,0.67261){\circle*{0.001}}
\put(0.84792,0.66640){\circle*{0.001}}
\put(0.85543,0.65979){\circle*{0.001}}
\put(0.86345,0.65271){\circle*{0.001}}

\put(0.86345,0.65271){\circle*{0.001}}
\put(0.87147,0.64527){\circle*{0.001}}
\put(0.87893,0.63762){\circle*{0.001}}
\put(0.88586,0.62979){\circle*{0.001}}
\put(0.89228,0.62180){\circle*{0.001}}
\put(0.89821,0.61369){\circle*{0.001}}
\put(0.90367,0.60548){\circle*{0.001}}
\put(0.90869,0.59719){\circle*{0.001}}
\put(0.91329,0.58885){\circle*{0.001}}
\put(0.91749,0.58047){\circle*{0.001}}
\put(0.92131,0.57208){\circle*{0.001}}
\put(0.92477,0.56368){\circle*{0.001}}
\put(0.92789,0.55529){\circle*{0.001}}
\put(0.93069,0.54693){\circle*{0.001}}
\put(0.93319,0.53860){\circle*{0.001}}
\put(0.93541,0.53033){\circle*{0.001}}
\put(0.93736,0.52211){\circle*{0.001}}
\put(0.93905,0.51395){\circle*{0.001}}
\put(0.94051,0.50587){\circle*{0.001}}
\put(0.94175,0.49787){\circle*{0.001}}
\put(0.94277,0.48996){\circle*{0.001}}
\put(0.94361,0.48214){\circle*{0.001}}
\put(0.94426,0.47441){\circle*{0.001}}
\put(0.94473,0.46678){\circle*{0.001}}
\put(0.94505,0.45925){\circle*{0.001}}
\put(0.94522,0.45183){\circle*{0.001}}
\put(0.94525,0.44451){\circle*{0.001}}
\put(0.94515,0.43730){\circle*{0.001}}
\put(0.94493,0.43020){\circle*{0.001}}
\put(0.94459,0.42321){\circle*{0.001}}
\put(0.94415,0.41633){\circle*{0.001}}
\put(0.94361,0.40956){\circle*{0.001}}
\put(0.94299,0.40290){\circle*{0.001}}
\put(0.94227,0.39635){\circle*{0.001}}
\put(0.94148,0.38991){\circle*{0.001}}
\put(0.94062,0.38359){\circle*{0.001}}
\put(0.93969,0.37737){\circle*{0.001}}
\put(0.93870,0.37125){\circle*{0.001}}
\put(0.93765,0.36525){\circle*{0.001}}
\put(0.93656,0.35935){\circle*{0.001}}
\put(0.93541,0.35355){\circle*{0.001}}

%3x3_sn=1%BP2
\put(0.80178,0.70710){\circle*{0.001}}
\put(0.80852,0.69871){\circle*{0.001}}
\put(0.81167,0.69004){\circle*{0.001}}
\put(0.81249,0.68140){\circle*{0.001}}
\put(0.81177,0.67296){\circle*{0.001}}
\put(0.81000,0.66483){\circle*{0.001}}
\put(0.80754,0.65705){\circle*{0.001}}
\put(0.80461,0.64963){\circle*{0.001}}
\put(0.80139,0.64259){\circle*{0.001}}
\put(0.79799,0.63592){\circle*{0.001}}
\put(0.79449,0.62960){\circle*{0.001}}
\put(0.79095,0.62361){\circle*{0.001}}
\put(0.78742,0.61794){\circle*{0.001}}
\put(0.78392,0.61258){\circle*{0.001}}
\put(0.78048,0.60749){\circle*{0.001}}
\put(0.77711,0.60267){\circle*{0.001}}
\put(0.77382,0.59809){\circle*{0.001}}
\put(0.77062,0.59375){\circle*{0.001}}
\put(0.76751,0.58962){\circle*{0.001}}
\put(0.76449,0.58569){\circle*{0.001}}
\put(0.76157,0.58195){\circle*{0.001}}
\put(0.75874,0.57839){\circle*{0.001}}
\put(0.75601,0.57499){\circle*{0.001}}
\put(0.75336,0.57174){\circle*{0.001}}
\put(0.75080,0.56864){\circle*{0.001}}
\put(0.74833,0.56568){\circle*{0.001}}

%3x3_sn=2%BP2
\put(2.30178,0.70710){\circle*{0.001}}
\put(2.29990,0.69911){\circle*{0.001}}
\put(2.29792,0.69129){\circle*{0.001}}
\put(2.29587,0.68366){\circle*{0.001}}
\put(2.29375,0.67623){\circle*{0.001}}
\put(2.29158,0.66901){\circle*{0.001}}
\put(2.28937,0.66199){\circle*{0.001}}
\put(2.28712,0.65519){\circle*{0.001}}
\put(2.28484,0.64860){\circle*{0.001}}
\put(2.28255,0.64222){\circle*{0.001}}
\put(2.28026,0.63605){\circle*{0.001}}
\put(2.27796,0.63009){\circle*{0.001}}
\put(2.27568,0.62433){\circle*{0.001}}
\put(2.27340,0.61878){\circle*{0.001}}
\put(2.27114,0.61342){\circle*{0.001}}
\put(2.26891,0.60825){\circle*{0.001}}
\put(2.26670,0.60326){\circle*{0.001}}
\put(2.26451,0.59845){\circle*{0.001}}
\put(2.26236,0.59382){\circle*{0.001}}
\put(2.26025,0.58935){\circle*{0.001}}
\put(2.25817,0.58504){\circle*{0.001}}
\put(2.25612,0.58088){\circle*{0.001}}
\put(2.25412,0.57687){\circle*{0.001}}
\put(2.25215,0.57301){\circle*{0.001}}
\put(2.25022,0.56928){\circle*{0.001}}
\put(2.24833,0.56568){\circle*{0.001}}
\put(0.65,0.4){$\omega_{3,2}$}
\put(0.97,0.3){$\omega_{3,1}$}
\put(2.15,0.4){$\omega_{3,2}$}
\put(2.47,0.3){$\omega_{3,1}$}
\put(2.31,0.95){$\omega_{2,1}$}
%\put(2.29,0.50){$W_{3,2}$}
%\put(0.82,0.50){$W_{3,2}$}

\end{picture}
\end{center}
\caption{
These pictures for biqutrit system show the regions for states, Schmidt number $3$ witnesses and Schmidt number $2$ witnesses, on
the two-dimensional planes through the states $\varrho_3$, $\varrho_1$, $\varrho_*$ and $\varrho_3$, $\varrho_2$, $\varrho_*$, respectively.
The isotropic states and the Choi matrices of Tomiyama maps are located
on the dotted line through the maximally entangled state $\varrho_3$ and the maximally mixed state $\varrho_*$.}
\end{figure}

We note that there exists no witness at all outside of the edge $\overline{\varrho_3 \varrho_i}$ for $i=1,2$
by Theorem \ref{boundary-2}, since every subspace of codimension $2$ has a product vector.
We also see that there is no witness at all outside of the edge $\overline{\varrho_3 \sigma_1}$
because the orthogonal of the range space is spanned by the product vector $|\xi_1\ran$.
We note that the one-dimensional subspace spanned by $|\xi_3\ran$ is exactly a $2$-entangled subspace, and so we apply
Theorem \ref{conc} to see that there exist Schmidt number $2$ and $3$ witnesses outside of the edge
$\overline{\varrho_i\sigma_i}$ for $i=1,2$.
Finally, we also see that there are Schmidt number 2 witnesses outside of the edge $\overline{\varrho_3\sigma_2}$,
because $|\xi_2\ran$ spans exactly a $1$-entangled subspace.

In order to figure out the regions for Schmidt number $2$ and $3$ witnesses, we compute
$\|\varrho_3\|_{S(k)}=\frac k3$ for $k=1,2,3$ and $\|\varrho_2\|_{S(k)}=\frac k2$ for $k=1,2$.
By (\ref{beta-}), we determine the boundary points
$\omega_{3,2} = {6 \over 5}\varrho_* -{1 \over 5}\varrho_3$ of $\blockpos_2$ and
$\omega_{3,1} = {3 \over 2}\varrho_* -{1 \over 2}\varrho_3$ of $\blockpos_1$ on $H_i$, and
$\omega_{2,1} = {9 \over 7}\varrho_* -{2 \over 7}\varrho_2$ of $\blockpos_1$ on $H_2$ additionally.
When we have two boundary points $W_1$ and $W_2$ of $\blockpos_k$, the line segment between
these two points lies on the boundary of $\blockpos_k$ if and only if there exists $|\xi\ran$ with $\SR|\xi\ran\le k$ satisfying
$\lan\xi|W_1|\xi\ran=0=\lan\xi|W_2|\xi\ran$.
With this principle, it is easy to see that the line segment $\overline{\varrho_i \omega_{3,k}}$
is on the boundary of $\blockpos_k$ on
$H_i$ for $i=1,2$ and $k=1,2$. One may also see that the line segments $\overline{\varrho_3 \omega_{2,1}}$ and $\overline{\omega_{2,1} \omega_{3,1}}$
are on the boundary of $\blockpos_1$ on $H_2$.
For the remaining boundaries, we take
$\varrho_i^\lambda=(1-\lambda)\varrho_3+\lambda\varrho_i$
and compute their operator $k$-norms by the formula (\ref{formu}).
Detailed arguments and computations will be provided in the Appendix.

%%%%%%%%%%%%%%%%%%%%%%%%%%%%%%%%%%%%%%%%%%%%%%%%%%%%%%%%%%%%%%%%%%%%%
%%%%%%%%%%%%%%%%%%%%%%%%%%%%%%%%%%%%%%%%%%%%%%%%%%%%%%%%%%%%%%%%%%%%%%%%%%%%%%%%%%%%%%%%%
%%%%%%%%%%%%%%%%%%%%%%%%%%%%%%%%%%%%%%%%%%%%%%%%%%%%%%%%%%%%%%%%%%%%%%%%%%%%%%%%%%%%%%%%%
%%%%%%%%%%%%%%%%%%%%%%%%%%%%%%%%%%%%%%%%%%%%%%%%%%%%%%%%%%%%%%%%%%%%%%%%%%%%%%%%%%%%%%%%%
%%%%%%%%%%%%%%%%%%%%%%%%%%%%%%%%%%%%%%%%%%%%%%%%%%%%%%%%%%%%%%%%%%%%%%%%%%%%%%%%%%%%%%%%%
%%%%%%%%%%%%%%%%%%%%%%%%%%%%%%%%%%%%%%%%%%%%%%%%%%%%%%%%%%%%%%%%%%%%%%%%%%%%%%%%%%%%%%%%%
\section{Conclusion}
%%%%%%%%%%%%%%%%%%%%%%%%%%%%%%%%%%%%%%%%%%%%%%%%%%%%%%%%%%%%%%%%%%%%%%%%%%%%%%%%%%
We have partitioned the region outside of the convex set $\mathcal D$ of all bipartite states, in order to
understand the global locations of Schmidt number witnesses. Existence of such witnesses outside of a face arising from a subspace
depends on the entanglement properties of its orthogonal complement. In particular, we have considered a one-parameter family $\{X_\lambda\}$ from
the maximally mixed state to arbitrary state, and determined the precise lower bound and upper bound for Schmidt
number witnesses on the family.

Determining the bounds of $\lambda$ for which $X_\lambda$ has a specific Schmidt number
must be a much more difficult problem, as we see in cases of Werner states \cite{Werner-1989} and isotropic states \cite{terhal-schmidt}.
It is also difficult to find Schmidt number witnesses which detect Schmidt numbers of
preassigned states. Our global geometry works for
the states on the face arising from $k$-entangled space \cite{SBL_2001}.
In order to find such Schmidt number witnesses for general cases, we
may have to examine \lq\lq local\rq\rq\ geometry around a specific face.
In {\sc FIG.\ 2}, we may need local geometry around $\sigma_i$
in order to determine Schmidt numbers around the edges connecting $\varrho_3$ and $\varrho_i$, for $i=1,2$.

%%%%%%%%%%%%%%%%%%%%%%%%%%%%%%%%%%%%%%%%%%%%%%%%%%%%%%%%%%%%%%%%%%%
%%%%%%%%%%%%%%%%%%%%%%%%%%%%%%%%%%%%%%%%%%%%%%%%%%%%%%%%%%%%%%%%%%%
%%%%%%%%%%%%%%%%%%%%%%%%%%%%%%%%%%%%%%%%%%%%%%%%%%%%%%%%%%%%%%%%%%%
%%%%%%%%%%%%%%%%%%%%%%%%%%%%%%%%%%%%%%%%%%%%%%%%%%%%%%%%%%%%%%%%%%%
%%%%%%%%%%%%%%%%%%%%%%%%%%%%%%%%%%%%%%%%%%%%%%%%%%%%%%%%%%%%%%%%%%%
\appendix

\section{An elementary proof of the identity for $\tau_k$}

During the proof of the formula (\ref{formu}), we used the following identity
\begin{equation}\label{formu-1}
\tau_k|\xi\ran=\sup\{|\lan\xi|\eta\ran|^2:\|\eta\|=1, \SR|\eta\ran\le k\},
\end{equation}
whose proof is given in  \cite[Theorem 3.3]{jonston_kribs}. The original proof
is quite involved using the Hardy-Littlewood-Polya theorem for doubly substochastic matrices.
Here we provide a simple elementary proof for the identity (\ref{formu-1}). We identify
an $m\times n$ matrix and a vector in $\mathbb C^m\ot\mathbb C^n$. Then, it suffices to prove the
following identity
\begin{equation}\label{cfvbngvghfb}
\tau_k(s)=\sup\{|\lan s|t\ran|^2: \|t\|_\HS=1,\ \rk t\le k\},
\end{equation}
for an $m\times n$ matrix $s$.

Take $t\in M_{m,n}$ with
$\|t\|_\HS=1$ and $\rk t\le k$. We denote by $Q$ the projection onto
$(\ker t)^\perp$. Then, we have
$$
\begin{aligned}
|\lan s|t\ran|^2 =|\tr(s^*t)|^2
= |\tr(Q s^*t)|^2
&\le ||Q s^*||_\HS^2 \\
&= \tr(Qs^*sQ) \le \sigma_1^2 + \sigma_2^2 +
\cdots + \sigma_k^2=\tau_k(s).
\end{aligned}
$$
On the other hand, we take the projection $Q$ onto the space spanned
by the eigenspaces of $s^*s$ associated with $\sigma_1^2,
\sigma_2^2, \cdots, \sigma_k^2$, and consider the rank $k$ operator
$t_0 = {sQ \over ||sQ||_\HS}$. Then, we have
$$
\begin{aligned}
|\tr(s^*t_0)|^2
&= {|\tr(s^*sQ)|^2 \over ||sQ||_\HS^2}=
{\tr(Qs^*sQ)^2 \over \tr(Qs^*sQ)}
& = \sigma_1^2 + \sigma_2^2 + \cdots
+ \sigma_k^2 =\tau_k(s),
\end{aligned}
$$
and so we get the required identity (\ref{cfvbngvghfb}).

%%%%%%%%%%%%%%%%%%%%%%%%%%%%%%%%%%%%%%%%%%%%%%%%%%%%%%%%%%%%%%%%%%%%%
\section{Boundaries of Schmidt number witnesses}
%%%%%%%%%%%%%%%%%%%%%%%%%%%%%%%%%%%%%%%%%%%%%%%%%%%%%%%%%%%%%%%%%%%%%

In this appendix, we determine the boundaries of the convex sets $\blockpos_1$ and $\blockpos_2$ on the planes
$H_1$ and $H_2$.

%%%%%%%%%%%%%%%%%%%%%%%%%%%%%%%%%%%%%%%%%%%%%%%%%%%%%%%%%%%%%%%%%%%
\subsection{Linear boundaries}
%%%%%%%%%%%%%%%%%%%%%%%%%%%%%%%%%%%%%%%%%%%%%%%%%%%%%%%%%%%%%%%%%%%

We first show that the line segment $\overline{\varrho_i \omega_{3,k}}$ is on the boundary of $\blockpos_k$ on
$H_i$ for $i,k=1,2$. To see this, we take a product vector $\lan \eta_1 |=(0,1,1) \otimes (0,1,1)$ and
$\lan\eta_2| = \lan00| +\lan11|$ with Schmidt rank $2$, to get the identities
$$
\lan \eta_k | \varrho_i | \eta_k \ran = 0 = \lan \eta_k | \omega_{3,k} | \eta_k \ran
$$
for $i,k=1,2$. Next, we take
$\lan \eta_3| = (1,{\rm i},0) \otimes (1,{\rm i},0)$ and $\lan\eta_4| = (1,1,0) \otimes (1,1,0)$,
to get the identities
$$
\lan \eta_3 | \varrho_3 | \eta_3 \ran = 0 = \lan \eta_3 | \omega_{2,1} |\eta_3 \ran
$$
together with
\begin{equation}\label{w21w31}
\lan \eta_4 | \omega_{2,1} | \eta_4 \ran = 0 = \lan \eta_4 | \omega_{3,1} |\eta_4 \ran.
\end{equation}
Therefore, we conclude that the line segments $\overline{\varrho_3 \omega_{2,1}}$ and $\overline{\omega_{2,1} \omega_{3,1}}$
are on the boundary of $\blockpos_1$ on $H_2$.

%%%%%%%%%%%%%%%%%%%%%%%%%%%%%%%%%%%%%%%%%%%%%%%%%%%%%%%%%%%%%%%%%%%
\subsection{Computation of operator $k$-norms of $\varrho_1^\lambda$}
%%%%%%%%%%%%%%%%%%%%%%%%%%%%%%%%%%%%%%%%%%%%%%%%%%%%%%%%%%%%%%%%%%%

Now, we proceed to compute the operator $k$-norms of
$$
\varrho_i^\lambda=(1-\lambda)\varrho_3+\lambda\varrho_i
$$
for $i=1,2$, in order to determine the remaining boundaries.
We use the formula (\ref{formu}).
For a unit vector $|\xi\ran$, we have
$$
(\varrho^\lambda_1)^{1/2} |\xi\ran = \sqrt{1-\lambda} \alpha_3 |\xi_3\ran
+ \sqrt{\lambda} \alpha_1 |\xi_1\ran, \qquad \alpha_i = \lan \xi_i|\xi\ran,
$$
where $\alpha_i$ satisfy $|\alpha_3|^2 + |\alpha_1|^2 \le 1$.
By the identification $\mathbb C^3 \otimes \mathbb C^3 = M_3$, it corresponds to the matrix
$$
A:=
\begin{pmatrix}
\sqrt{1-\lambda \over 3} \alpha_3 & \sqrt{\lambda} \alpha_1 &  \\
 & \sqrt{1-\lambda \over 3} \alpha_3 & \\
 & & \sqrt{1-\lambda \over 3} \alpha_3
\end{pmatrix}
\in M_3,
$$
and the eigenvalues of $A^*A$ are given by
\begin{equation}\label{eigen1}
\mu:={1 -\lambda \over 3} |\alpha_3|^2, \qquad
\mu_{\pm} := \pm |\alpha_1| \sqrt{\lambda} \sqrt{ {1-\lambda \over 3} |\alpha_3|^2 + {\lambda \over 4}|\alpha_1|^2}
+ {1-\lambda \over 3}|\alpha_3|^2 + {\lambda \over 2} |\alpha_1|^2.
\end{equation}
Since their sum is $(1-\lambda) |\alpha_3|^2 + \lambda |\alpha_1|^2$,
we have
$$
||\varrho^\lambda_1||_{S(3)} =
\begin{cases} 1 -\lambda, &  0 \le \lambda \le {1 \over 2} \\ \lambda, & {1 \over 2} \le \lambda \le 1\end{cases}.
$$
Alternatively, it follows immediately from
$||\varrho^\lambda_1|| = || (1-\lambda)\varrho_3+\lambda \varrho_1|| = \max\{1-\lambda, \lambda\}$.

Among three eigenvalues in (\ref{eigen1}), the number $\mu_+$ is greatest.
We suppose that $|\alpha_3|^2+|\alpha_1|^2=1$ and take $t=|\alpha_1|$.
Let us maximize the function
$$
f(t) = t \sqrt{\lambda} \sqrt{ {1-\lambda \over 3} (1-t^2) + {\lambda \over 4}t^2}
+ {1-\lambda \over 3}(1-t^2) + {\lambda \over 2} t^2.
$$
When ${4 \over 7} \le \lambda \le 1$, the only critical point is $-1$ and $f$ is increasing on $[0,1]$.
Otherwise, $f$ has three critical points
$$
-1,\quad \pm \sqrt{3\lambda \over 4-7\lambda}.
$$
We have
$0 \le \sqrt{3\lambda \over 4-7\lambda} \le 1$ if and only if $0 \le \lambda \le {2 \over 5}$. In this case, $f$
takes the maximum
$f\left(\sqrt{3\lambda \over 4-7\lambda}\right)
= {4 \over 3} {(1-\lambda)^2 \over 4-7\lambda}$.
When ${2 \over 5} \le \lambda \le 1$, $f$ is increasing on $[0,1]$ with a maximum
$f(1)=\lambda$.
Therefore, we see that
$$
||\varrho^\lambda_1||_{S(1)} =
\begin{cases}{4 \over 3} {(1-\lambda)^2 \over 4-7\lambda}, &  0 \le \lambda \le {2 \over 5} \\
\lambda, & {2 \over 5} \le \lambda \le 1\end{cases}.
$$

\begin{figure}
\includegraphics[scale=0.5]{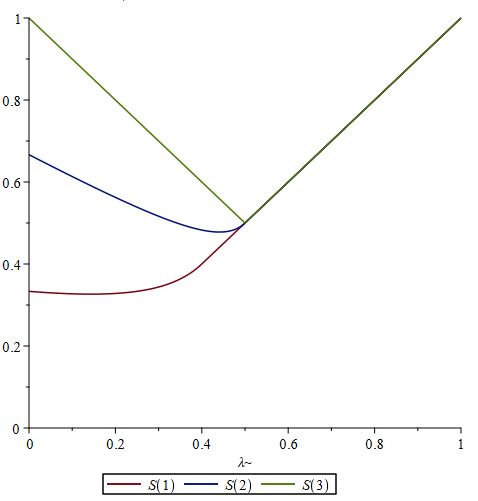}
\caption{The graphs of $||\varrho^\lambda_1||_{S(k)}$ with $k=1,2,3$.}
\label{SR3SR1}
\end{figure}

We proceed to compute $||\varrho^\lambda_1||_{S(2)}$.
The sum of $\mu_+$ and $\mu_-$ is
${2(1-\lambda) \over 3}|\alpha_3|^2 + \lambda |\alpha_1|^2$,
whose maximum is given by
$$
M_\lambda = \begin{cases}{2(1-\lambda) \over 3}, &  0 \le \lambda \le {2 \over 5} \\
\lambda, & {2 \over 5} \le \lambda \le 1\end{cases}.
$$
On the other hand, the sum of $\mu$ and $\mu_+$ is given by
$$
|\alpha_1| \sqrt{\lambda} \sqrt{ {1-\lambda \over 3} |\alpha_3|^2 + {\lambda \over 4}|\alpha_1|^2}
+ {2(1-\lambda) \over 3}|\alpha_3|^2 + {\lambda \over 2} |\alpha_1|^2,
$$
and so we maximize the function
$$
g(t) = t \sqrt{\lambda} \sqrt{ {1-\lambda \over 3} (1-t^2) + {\lambda \over 4}t^2}
+ {2(1-\lambda) \over 3}(1-t^2) + {\lambda \over 2} t^2.
$$
When ${4 \over 7} \le \lambda \le 1$, $g$ increases and its maximum is $g(1)=\lambda$. Otherwise, it has four critical points
$$
t_{\pm,\pm} = \pm \sqrt{14 \lambda^2 -22\lambda+8 \pm \sqrt{343\lambda^4-931\lambda^3+924\lambda^2-400\lambda+64}
\over 49\lambda^2-56\lambda+16}.
$$
Among them, only $t_{+,-}$ for $0 \le \lambda \le {1 \over 2}$ lies in $[0,1]$.
We have
$g(t_{+,-}) = {1 \over 3} \left( 2\sqrt{1-\lambda \over 4-7\lambda} +1\right)(1-\lambda)$.
Therefore, the maximum of $g$ is given by
$$
N_\lambda = \begin{cases} {1 \over 3} \left( 2\sqrt{1-\lambda \over 4-7\lambda} +1\right)(1-\lambda), &  0 \le \lambda \le {1 \over 2} \\
\lambda, & {1 \over 2} \le \lambda \le 1\end{cases}.
$$
Since $N_\lambda \ge M_\lambda$ for all $0 \le \lambda \le 1$, we have
$$
||\varrho^\lambda_1||_{S(2)} = \begin{cases} {1 \over 3} \left( 2\sqrt{1-\lambda \over 4-7\lambda} +1\right)(1-\lambda), &  0 \le \lambda \le {1 \over 2} \\
\lambda, & {1 \over 2} \le \lambda \le 1\end{cases}.
$$

See {\sc FIG.\ \ref{SR3SR1}} for the graphs of $||\varrho^\lambda_1||_{S(k)}$ with $k=1,2,3$.
When ${1 \over 2} \le \lambda \le 1$, the eigenspace corresponding to the largest eigenvalue is ${\rm span} |\xi_1\ran$
which is not 1-entangled. The graphs on $[1/2,1]$ illustrate Theorem \ref{boundary-}.

%%%%%%%%%%%%%%%%%%%%%%%%%%%%%%%%%%%%%%%%%%%%%%%%%%%%%%%%%%%%%%%%%%%
\subsection{Computation of operator $k$-norms of $\varrho_2^\lambda$}
%%%%%%%%%%%%%%%%%%%%%%%%%%%%%%%%%%%%%%%%%%%%%%%%%%%%%%%%%%%%%%%%%%%

For a unit vector $|\xi\ran$, we have
$$
(\varrho^\lambda_2)^{1/2} |\xi\ran = \sqrt{1-\lambda} \alpha_3 |\xi_3\ran
+ \sqrt{\lambda} \alpha_2 |\xi_2\ran, \qquad \alpha_i = \lan \xi_i|\xi\ran,
$$
where $\alpha_i$ satisfy $|\alpha_3|^2 + |\alpha_2|^2 \le 1$.
By the identification $\mathbb C^3 \otimes \mathbb C^3 = M_3$, it corresponds to the matrix
$$
B:=
\begin{pmatrix}
\sqrt{1-\lambda \over 3} \alpha_3 & \sqrt{\lambda \over 2} \alpha_2 & \\
\sqrt{\lambda \over 2} \alpha_2 & \sqrt{1-\lambda \over 3} \alpha_3 & \\
 & & \sqrt{1-\lambda \over 3} \alpha_3
\end{pmatrix}
\in M_3,
$$
and the eigenvalues of $B^*B$ are given by
\begin{equation}\label{eigen2}
\nu:={1 -\lambda \over 3} |\alpha_3|^2, \qquad
\nu_{\pm}:={1 -\lambda \over 3} |\alpha_3|^2
+ {\lambda \over 2} |\alpha_2|^2 \pm \sqrt{\lambda (1-\lambda) \over 6} |\alpha_3 \bar \alpha_2 + \bar \alpha_3 \alpha_2|.
\end{equation}

Since their sum is $(1-\lambda) |\alpha_3|^2 + \lambda |\alpha_2|^2$,
we have
$$
||\varrho^\lambda_2||_{S(3)} =
\begin{cases} 1 -\lambda, &  0 \le \lambda \le {1 \over 2} \\ \lambda, & {1 \over 2} \le \lambda \le 1\end{cases}.
$$
Alternatively, it may follow immediately from
$||\varrho_2^\lambda|| = || (1-\lambda)\varrho_3+\lambda \varrho_2|| = \max\{1-\lambda, \lambda\}$.

Among three eigenvalues in (\ref{eigen2}), the number $\nu_+$ is greatest and satisfies
$$
\begin{aligned}
\nu_+ \le &{1 -\lambda \over 3} |\alpha_3|^2 + {\lambda \over 2} |\alpha_2|^2 + 2\sqrt{\lambda (1-\lambda) \over 6} |\alpha_3 \alpha_2| \\
= & \left(\sqrt{1 -\lambda \over 3} |\alpha_3| + \sqrt{\lambda \over 2} |\alpha_2|\right)^2 \\
\le & \left({1 -\lambda \over 3} + {\lambda \over 2}\right)(|\alpha_3|^2+|\alpha_2|^2) \\
\le &  {2+\lambda \over 6}.
\end{aligned}
$$
If $(\alpha_3,\alpha_2)$ is a unit vector of positive scalar multiple of $(\sqrt{1 -\lambda \over 3},\sqrt{\lambda \over 2})$,
then the equality holds.
Thus, we have
$$
||\varrho_2^\lambda||_{S(1)} = {2+\lambda \over 6}.
$$

The maximum of the sum of eigenvalues $\nu_+$ and $\nu_-$ in (\ref{eigen2}) is also given  by
$$
M_\lambda = \begin{cases}{2(1-\lambda) \over 3} & , 0 \le \lambda \le {2 \over 5} \\  \lambda &, {2 \over 5} \le \lambda \le 1\end{cases}.
$$
On the other hand, the sum of $\nu$ and $\nu_+$ satisfies
$$
\begin{aligned}
\nu + \nu_+
\le & {2(1 -\lambda) \over 3} |\alpha_3|^2 + 2 \sqrt{\lambda (1-\lambda) \over 6} |\alpha_3\alpha_2| + {\lambda \over 2} |\alpha_2|^2 \\
=& \begin{pmatrix} |\alpha_3| & |\alpha_2| \end{pmatrix}
\begin{pmatrix} {2(1 -\lambda) \over 3} &  \sqrt{\lambda (1-\lambda) \over 6} \\
\sqrt{\lambda (1-\lambda) \over 6} & {\lambda \over 2} \end{pmatrix}
\begin{pmatrix} |\alpha_3| \\ |\alpha_2| \end{pmatrix},
\end{aligned}
$$
and its maximum is achieved as the bigger eigenvalue,
$$
L_\lambda:={4-\lambda +\sqrt{25\lambda^2 - 32 \lambda +16} \over 12},
$$
of the middle matrix, when $(\alpha_3,\alpha_2)$ is a unit eigenvector with positive entries.
One may verify that $L_\lambda \ge {2(1 -\lambda) \over 3}$ for all $0 \le \lambda \le 1$, and
$L_\lambda \ge \lambda$ if and only if $0 \le \lambda \le {1 \over 2}$.
Therefore, we see that
$$
||\varrho_2^\lambda||_{S(2)} =
\begin{cases}{4-\lambda +\sqrt{25\lambda^2 - 32 \lambda +16} \over 12}, &  0 \le \lambda \le {1 \over 2} \\
\lambda, & {1 \over 2} \le \lambda \le 1\end{cases}.
$$

\begin{figure}
\includegraphics[scale=0.5]{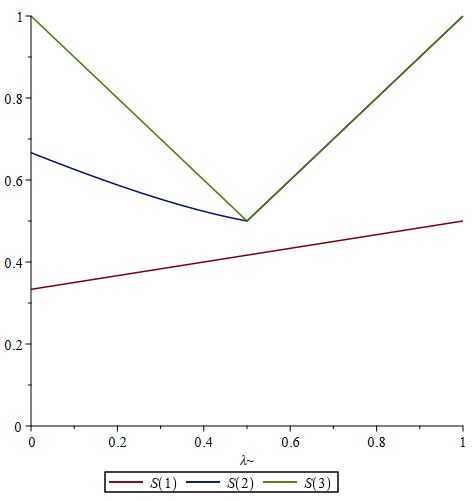}
\caption{The graphs of $||\varrho^\lambda_2||_{S(k)}$ with $k=1,2,3$.}
\label{SR3SR2}
\end{figure}

See {\sc FIG.\ \ref{SR3SR2}} for the graphs of $||\varrho^\lambda_2||_{S(k)}$ with $k=1,2,3$.
When ${1 \over 2} \le \lambda \le 1$, the eigenspace corresponding to the largest eigenvalue is
${\rm span} |\xi_2\ran$ which is exactly 1-entangled. Note that the graphs on the interval $[1/2,1]$ illustrate Theorem \ref{boundary-},
and the linearity of $||\varrho^\lambda_2||_{S(1)}$ is consistent with (\ref{w21w31}).

%%%%%%%%%%%%%%%%%%%%%%%%%%%%%%%%%%%%%%%%%%%%%%%%%%%%%%%%%%%%%%%%%%%%%
%%%%%%%%%%%%%%%%%%%%%%%%%%%%%%%%%%%%%%%%%%%%%%%%%%%%%%%%%%%%%%%%%%%%%%%%%%%%%%%%%%%%%%%%%
%%%%%%%%%%%%%%%%%%%%%%%%%%%%%%%%%%%%%%%%%%%%%%%%%%%%%%%%%%%%%%%%%%%%%%%%%%%%%%%%%%%%%%%%%
%%%%%%%%%%%%%%%%%%%%%%%%%%%%%%%%%%%%%%%%%%%%%%%%%%%%%%%%%%%%%%%%%%%%%%%%%%%%%%%%%%%%%%%%%
%%%%%%%%%%%%%%%%%%%%%%%%%%%%%%%%%%%%%%%%%%%%%%%%%%%%%%%%%%%%%%%%%%%%%%%%%%%%%%%%%%%%%%%%%
%%%%%%%%%%%%%%%%%%%%%%%%%%%%%%%%%%%%%%%%%%%%%%%%%%%%%%%%%%%%%%%%%%%%%%%%%%%%%%%%%%%%%%%%%
%%%%%%%%%%%%%%%%%%%%%%%%%%%%%%%%%%%%%%%%%%%%%%%%%%%%%%%%%%%%%%%%%%%%%%%%

\end{document}